\documentclass[useAMS, usenatbib]{mnras}
\pdfoutput=1
\usepackage{amssymb}
\usepackage{gensymb}
\usepackage{amsmath}
\newcommand{\photo}{\textsc{photo}}

\def\msun{\hbox{M$_\odot$}}
\def\mstar{\hbox{M$_\star$}}

\def\one{\,{\sc i}}             % for producing Na I as Na\one\ etc.
\def\two{\,{\sc ii}}

\usepackage{graphicx}
\title[GAMA: detection of low-SB galaxies]{Galaxy And Mass Assembly (GAMA): detection of low-surface-brightness
  galaxies from SDSS data}
\author[R. P. Williams et al.]{Richard P. Williams,$^{1}$ I. K. Baldry,$^{1}$ L. S. Kelvin,$^{1}$ P. A. James,$^{1}$ S. P.~Driver,$^{2,3}$\newauthor M. Prescott,$^{4}$ S.~Brough,$^{5}$ M. J. I.~Brown,$^{6}$ L. J. M. Davies,$^{2}$ B. W. Holwerda,$^{7}$\newauthor J. Liske,$^{8}$ P.~Norberg,$^{9}$ A. J. Moffett,$^{2}$ A. H. Wright$^{2}$ \\
$^{1}$Astrophysics Research Institute, Liverpool John Moores University, IC2, Liverpool Science Park, 146 Brownlow Hill, L3 5RF, UK\\
$^{2}$International Centre for Radio Astronomy Research (ICRAR), 
University of Western Australia, Crawley, WA 6009, Australia\\ 
$^{3}$Scottish Universities' Physics Alliance (SUPA), School of Physics and 
Astronomy, University of St Andrews, KY16 9SS, UK\\
$^{4}$Department of Physics and Astronomy, University of the Western Cape, Private Bag X17, Bellville 7535, South Africa\\
$^{5}$Australian Astronomical Observatory, PO Box 915, North Ryde, NSW 1670, Australia\\
$^{6}$School of Physics, Monash University, Clayton, Victoria 3800, Australia\\
$^{7}$University of Leiden, Sterrenwacht Leiden, Niels Bohrweg 2, NL-2333 CA Leiden, The Netherlands\\
$^{8}$Hamburger Sternwarte, Universit{\''at} Hamburg, Gojenbergsweg 112, 21029 Hamburg, Germany\\
$^{9}$ICC \& CEA, Department of Physics, Durham University, South Road, DH1 3LE, UK\\ 
}
\date{Accepted by MNRAS, 2016 August 29}
\pagerange{\pageref{firstpage}--\pageref{lastpage}}
\pubyear{2016}
\begin{document}
\maketitle
\label{firstpage}
\begin{abstract}

We report on a search for new low-surface-brightness galaxies (LSBGs) using Sloan Digital Sky Survey (SDSS) data within the GAMA equatorial fields. The search method consisted of masking objects detected with SDSS \photo, combining {\it gri} images weighted to maximise the expected signal-to-noise ratio (SNR), and smoothing the images. The processed images were then run through a detection algorithm that finds all pixels above a set threshold and groups them based on their proximity to one another. The list of detections was cleaned of contaminants such as diffraction spikes and the faint wings of masked objects. From these, selecting potentially the brightest in terms of total flux, a list of 343 LSBGs was produced having been confirmed using VISTA Kilo-degree Infrared Galaxy Survey (VIKING) imaging. The photometry of this sample was refined using the deeper VIKING $Z$ band as the aperture-defining band. 
Measuring their $g-i$ and $J-K$ colours shows that most are consistent with being at redshifts less than 0.2. The photometry is carried out using an \textsc{auto} aperture for each detection giving surface brightnesses of $\mu_{r} \ga 25$\,mag arcsec$^{-2}$ and magnitudes of $r > 19.8$\,mag. None of these galaxies are bright enough to be within the GAMA main survey limit
but could be part of future deeper surveys to measure the low-mass end of the galaxy stellar mass function. 
\end{abstract}

\section{Introduction}
\label{sec:intro}

%%\vspace{20mm}

Most galaxy surveys to date have been limited by a combination of apparent magnitude and surface-brightness (SB) constraints. This has led to an over-representation of luminous high-SB galaxies compared to a complete volume-limited sample \citep{Cross+02}. Generally, flux-limited samples have been used to construct our picture of galaxy types, e.g., the Hubble Tuning Fork \citep{Hubble+1926}. However, the majority of galaxies are, in fact, low-luminosity or low-mass `dwarfs' \citep{Binggeli+88,Karachentsev+04,Driver+05,Baldry+2012}. Whilst many are late-type spirals, they often do not fit neatly into the `tuning fork'. Dwarf galaxies have, for example, been classified as: irregulars \citep{Hubble+1926, DeVauc+1959}, dwarf ellipticals \citep{Shapley+38}, dwarf spheroidals \citep{DeVauc+1968}, blue compact dwarfs \citep{Zwicky+71}, little blue spheroids \citep{Kelvin+14}, blue diffuse dwarfs \citep{James+15}, and ultra diffuse galaxies \citep{vanDokkum+15}.\footnote{Note we do not include ultra-compact dwarfs \citep[UCDs;][]{Phillipps+01} in this classification list. They typically have half-light radii closer to the values of globular clusters \citep{Gilmore+07}, and are most likely the central star-cluster remnants of larger galaxies \citep{Jennings+15, Janz+16}. As \cite{Kissler+04} argued, they are ``neither dwarf galaxies nor ultra compact''.}

$\Lambda$ Cold Dark Matter ($\Lambda$CDM) simulations have been used to make predictions for the number density of low-mass galaxies. When compared to observations these simulations show a discrepancy, known as the substructure problem \citep{Moore+99}, which can be characterised in two distinct ways. The first is the so-called missing satellite problem: the deficiency of the number of observed satellites, around the Milky Way in particular, compared to the number of sub-halos predicted by models \citep{Klypin+99, Moore+99}. The second deals with the discrepancy between the predicted number of halos and observed galaxies on a cosmological scale \citep[e.g.,][]{Peebles+01}. 

There has been progress towards reducing the discrepancy between simulations and observations in the Local Group with the discovery of many faint dwarf galaxies around the Milky Way \citep{Gilmore+07,Irwin+07,Walsh+07,Belokurov+10} and M31 \citep{Ibata+07,Martin+09,Richardson+11,Martin+132}. Furthermore, the number of galaxies in simulations can be reduced by, for example, changing cold dark matter to warm dark matter \citep{Xi+13,Lapi+15}, or by suppression of dwarf galaxy formation from a photo-ionizing background \citep{Benson+02_2,Somerville02}.

Unlike their satellite counterparts, which have a more complex and turbulent formation history, field dwarf galaxies form and evolve in isolation. Despite processes such as supernova feedback \citep{Ferrara+2000} and heating from the cosmic ionising background radiation \citep{Hoeft+06}, they generally have a larger cool gas fraction and higher star formation rate than those dwarfs that are gravitationally bound to a larger system. Instead of being stripped away, most of their gas can cool back into the system \citep{Rosenbaum+09}. Thus, the simulations and observations of low-mass field galaxies test a different regime to satellite galaxies. 

There is currently a significant difference in the number density of low-mass systems ($10^{6.5}\msun < \mstar < 10^{7} \msun$) between observations and simulations. For instance \cite{Guo+11}, through the use of simulations, predicted a number density of 0.1 Mpc$^{-3}$ dex$^{-1}$. Currently the best observations put that number density 
at $\sim$ 0.02 Mpc$^{-3}$ dex$^{-1}$ within this mass range, from \cite{Baldry+2012} using the Galaxy And Mass Assembly \citep[GAMA;][]{Driver+09b} survey. Therefore, observations must push to deeper magnitudes, and lower masses, in order to test whether observational SB limits are the reason, or part of the reason, for the discrepancy.

The detection of faint low-mass galaxies is challenging, dwarf systems have an intrinsically lower surface brightness than their higher-mass counterparts and so are more difficult to detect against the sky \citep{Disney+76,Disney+83,Kormendy+85,Baldry+08}. A typical definition in the literature for low-surface-brightness galaxies (LSBGs) is to have a central surface brightness of $\mu_{B} \ga 23$ mag arcsec$^{-2}$ \citep{McGaugh+96,Impey+1997}. This surface brightness makes them difficult to detect against the sky and leads to detection biases \citep{Disney+76}. 
Finding LSBGs is thus key to accounting for, and characterising, the dwarf galaxy populations of both satellites and isolated galaxies. 
A full accounting is needed to comprehensively test models of galaxy formation that include low-mass galaxies.

\subsection{Searches for field dwarf galaxies}

There are different environments to consider when searching for LSBGs and these environments can be broadly defined as: 
(i) nearby satellite galaxies within the Local Group (e.g.\ \citealt{Koposov+08,Walsh+09}); 
(ii) satellites in external groups and clusters (e.g.\ \citealt{Davies+16,vanDokkum+15}); and 
(iii) field galaxies away from luminous galaxies and clusters, or within a random cosmological volume. Compared to the Local Group, where stars can be resolved, and around luminous galaxies and in clusters, where deep imaging is more easily done and membership is more easily assigned, finding LSBGs in the field is more problematic. A large area of the sky needs to be covered in order to obtain a cosmologically representative sample. This means that a lower depth is obtained in the imaging compared with targeted cluster surveys given the same amount of observing time. In addition, redshifts need to be obtained for galaxies in order to assign distances \citep{BlantonLowLum+05,Geller+12}. 

To improve detection of these systems, specialised algorithms can be used to find LSBGs in images from wide surveys such as the Sloan Digital Sky Survey \citep[SDSS,][]{York+2000}. \cite{Kniazev+04} used SDSS data to search for galaxies of large angular size recovering most of the LSBGs from the \citet{Impey+96} catalogue. 
Fainter features can be found by coadding images paying careful attention to sky subtraction \citep{Fliri+16}; 
and/or known galaxies can be masked out, 
meaning specialised algorithms can be applied to the images to search for fainter light from LSBGs that were not initially detected \citep{Scaramella+09}. 
Similar techniques, including smoothing of masked images, can be used to search for low-SB tidal features \citep{Miskolczi+11}. 
\cite{James+15} used a search of the SDSS data to search for galaxies with similar morphology to Leo~P \citep{Giovanelli+13}, 
which has embedded H\two\ regions within a blue diffuse galaxy, and were able to detect $\sim 100$ of these sources.

Star-forming dwarf galaxies dominate the field dwarf population \citep{Geha+12}. 
Therefore they can be detected using radio H\one\ surveys, such as the Arecibo Legacy Fast ALFA survey \citep[ALFALFA,][]{Giovanelli+05}, 
because they typically have high H\one\ to stellar mass ratios \citep{Baldry+08,Huang+12}. 
The searching of optical images is then eased considerably, by knowing where to look, for example: 
\cite{Trachternach+06} and \cite{Du+15} have confirmed many hundreds of new LSBGs based on their H\one\ detections, mostly in the field;  
\cite{Sand+15} confirmed five new blue diffuse dwarf galaxies within 10 Mpc, associated with `high-velocity clouds' found in ALFALFA data; and 
\cite{Tollerud+15}, using a blind H\one\ survey \citep[GALFA-H\one, ][]{Peek+11}, were able to detect two more faint diffuse galaxies, again within 10 Mpc. 

\subsection{Aims of this analysis}
\label{sec:aims}

The galaxy stellar mass function (GSMF) is a fundamental tool used in studying the demographics of galaxies \citep{Bell+03,Baldry+08,Muzzin+13}. It describes the number density of galaxies as a function of their mass within a volume of the Universe. The GAMA survey team has accurately described the GSMF down to $\mstar = 10^8 \msun$. The current incarnation, however, is likely incomplete at masses below this due to SB limits \citep{Baldry+2012}. As such, in order to push below this limit it is important to carry out a search of the SDSS DR7 data within the GAMA fields. SDSS data have been chosen for this work as this survey has already demonstrated its suitability for finding low-SB systems \citep{Kniazev+04}.

The GAMA survey has made significant progress towards uncovering and classifying the dwarf population. For instance, \cite{Baldry+2012} showed that the most common type of galaxy in the Universe is likely star-forming dwarf galaxies rather than passive. \cite{Kelvin+14,Kelvin+14b} measured the contribution of `little blue spheroids', Sd spirals and irregulars to the low-mass and low-luminosity number densities. \cite{Mahajan+15} showed that star-forming dwarf galaxies formed a unimodal population using various photometric
and derived properties. 
However, progress still needs to be made into the search for, and detection of, LSBGs within this survey to work towards completing the census of galaxies.

This search is complicated, and the method employed depends on the type of data that is provided and the nature of the objects being searched for. The distance range desired for the detection of LSBGs in this paper is around 10 to 100\,Mpc, which places them beyond the range of the local group and volume \citep{McConnachie+12}. The volume
out to 100\,Mpc over the GAMA equatorial fields is 18\,000\,Mpc$^3$. This is more cosmologically representative than studies in the local volume ($<10$\,Mpc) because 
of the larger volume and longer sight lines that cut through filaments and void-like regions. 
A specialised detection algorithm was developed to detect LSBGs, which are difficult to detect because of sky noise and artefacts. 

This paper deals with the method for the creation and implementation of such a search algorithm to find these LSBGs within SDSS imaging, with comparisons to VISTA Kilo-degree Infrared Galaxy Survey \citep[VIKING,][]{Edge+13} imaging. This is being used to confirm or deny a detection as the VIKING $Z$ band is $\sim 1$ magnitude deeper than SDSS $r$ band when compared to an average SED for a low-redshift galaxy \citep{Driver+12}. In future, we plan to focus on finding similar objects in VIKING and, eventually, Kilo-Degree Survey (KiDS; \citealt{kuijken+11}) images which have deeper limits in surface brightness than SDSS. Initially this can be done using a standard method (e.g.\ Source Extractor), before using this information to create masks to then apply the methods described and tested in this paper.

The outline of the paper is as follows. In \S\,\ref{sec:surveys}, the different survey data are described; \S\,\ref{sec:ImRed} deals with the development and implementation of the image processing code; \S\,\ref{sec:DetAl} describes the algorithm used to search the images to find the LSBGs hidden within; and \S\,\ref{sec:Results} presents the results and catalogue. Summary and conclusions are presented in \S\,\ref{sec:sum+con}.

\section{Surveys}
\label{sec:surveys}

The newly detected objects presented in this work were discovered using SDSS $gri$ imaging within the limits of the three GAMA equatorial regions, G09, G12, and G15. For confirmation that these detections are galaxies, further visual confirmation was required. This was achieved through the use of deep VIKING $Z$-band data. The technical details of these surveys are described below.

\subsection{GAMA}
\label{sec:gama}

The GAMA survey is a wide-field spectroscopic survey that was undertaken to study cosmology, galaxy structure and galaxy evolution at low redshift \citep{Driver+11,Liske+15}. Redshifts of galaxies were obtained using the AAOmega spectrograph on the Anglo Australian Telescope \citep{Sharp+06,Hopkins+13}. Spectra have been obtained for 238\,000 objects in five survey regions (G02, G09, G12, G15, and G23, which are fields centred at RA 2h, 9h, 12h, 14.5h, and 23h), with a limiting magnitude of $r < 19.8$ for the main survey over four of the fields. The total survey area is 286 square degrees.   Independent imaging has been compiled from several other surveys whose footprints fall on the GAMA regions, covering wavelengths from the far ultraviolet to the radio \citep{Driver+16}. 

This paper primarily uses data from the three equatorial regions G09, G12, and G15, due to coverage of these areas by SDSS and VIKING. 
The galaxy stellar masses used in this paper were calculated using the method outlined in \cite{Taylor+11} updated to include 
VIKING $Z$- and $Y$-band data, in addition to SDSS data, in the fitting procedure. 
For these, a cosmology with $H_{0} = 70\,\mathrm{km\,s^{-1}\,Mpc^{-1}}$, $\Omega_{m,0} = 0.3$ and $\Omega_{\Lambda,0} = 0.7$ was assumed.
Note these stellar masses and redshifts from the GAMA survey are used to define comparison samples, 
none of the newly detected LSBGs described in this paper have spectroscopic redshifts. 

\subsection{Sloan Digital Sky Survey (SDSS)}
\label{sec:sdss}

The Sloan Digital Sky Survey (SDSS; \citealt{York+2000}) has observed over 10\,000 square degrees of sky. The imaging was done using thirty 2048$\,\times\,$2048 CCDs and five filters $u$, $g$, $r$, $i$ and $z$ \citep{Fukugita+1996}. Almost all of its standard imaging data were released in Data Release 7 \citep{Abazajian+09}. 
This imaging was taken in `drift scan' mode. Each part of the sky was exposed for about 55 seconds as the sky moves across the detector, which was read out at the same rate. This created long strips of images in the scan direction. These strips, one for each detector, were subsequently processed through the SDSS \photo\ pipeline \citep{Stoughton+02} and divided into fields along the scan direction for convenience of use. 

The data used in this work are the corrected images in the DR7 database. All information stored in the images is presented in counts, which can be converted into the AB magnitude system by applying equations supplied by SDSS. These images are supplemented by various masks for each filter. A code supplied by SDSS is used to extract the type of mask desired by the user 
\citep{Stoughton+02}.\footnote{The corrected image files start with 
prefix fpC. The mask files start with prefix fpM, and were read using 
\textsc{readAtlasImages-v5\_4\_11}.}

\subsection{VISTA VIKING}
\label{sec:vik}

The Visible and Infrared Survey Telescope for Astronomy (VISTA, \citealt{Emerson+10}) is a 4.1-metre wide-field telescope. It is located on Paranal Observatory in Chile. One of the surveys this telescope is carrying out is the VIKING survey. This survey covers $1400$ deg$^2$ in the near infrared $Z$, $Y$, $J$, $H$, and $K$ bands. The $Z$ band, in particular, was used as a check for the detections due to its improved depth of imaging over the SDSS bands. Notably the VISTA $Z$ band was taken in dark time, unlike typical $z$-band observations that are `competing' with visible bands.

\section{Image Processing}  
\label{sec:ImRed}

The reduced SDSS images were processed in order to specifically search
for LSBGs that had been missed by the SDSS pipeline. 
The image processing can be separated into five distinct phases: 
(i) masking of image fields; 
(ii) alignment of images;
(iii) weighting to maximise the expected SNR; 
(iv) coadding of $g$, $r$, and $i$ image fields; and
(v) smoothing of the final image products.
These processes are described in the following paragraphs.

\begin{figure*}
\centering
\includegraphics[scale=0.55]{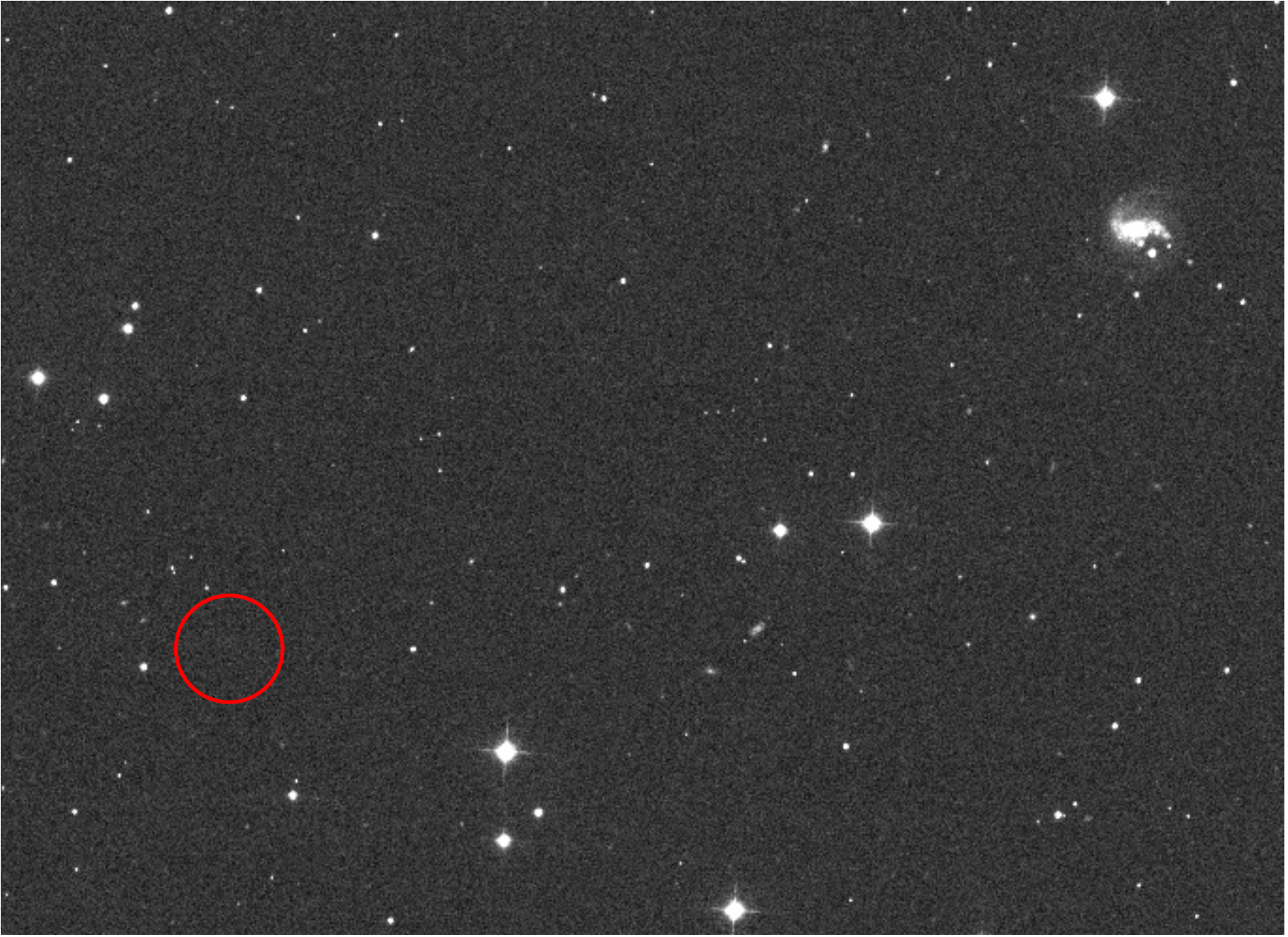}
\includegraphics[scale=0.55]{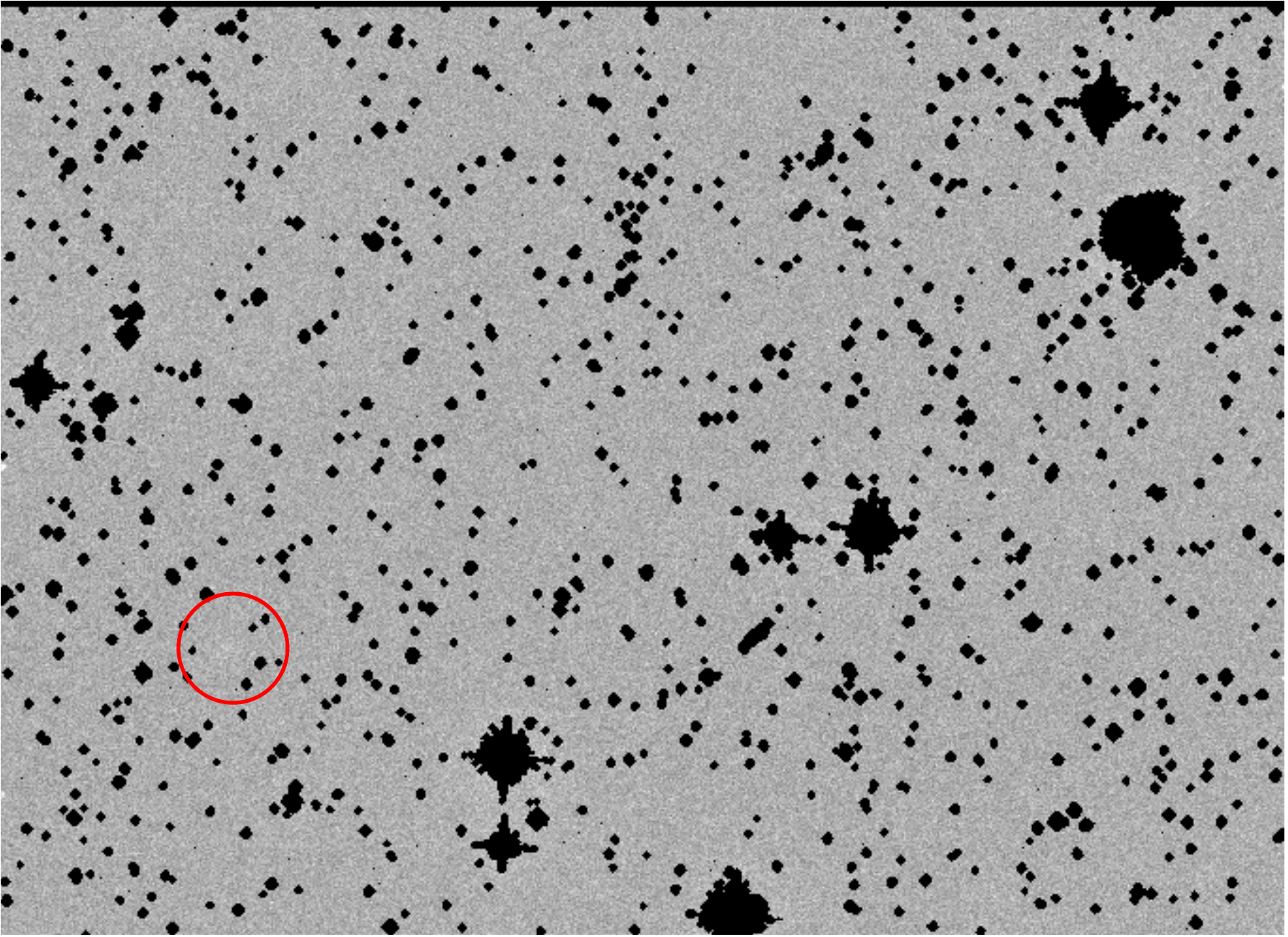}
\includegraphics[scale=0.55]{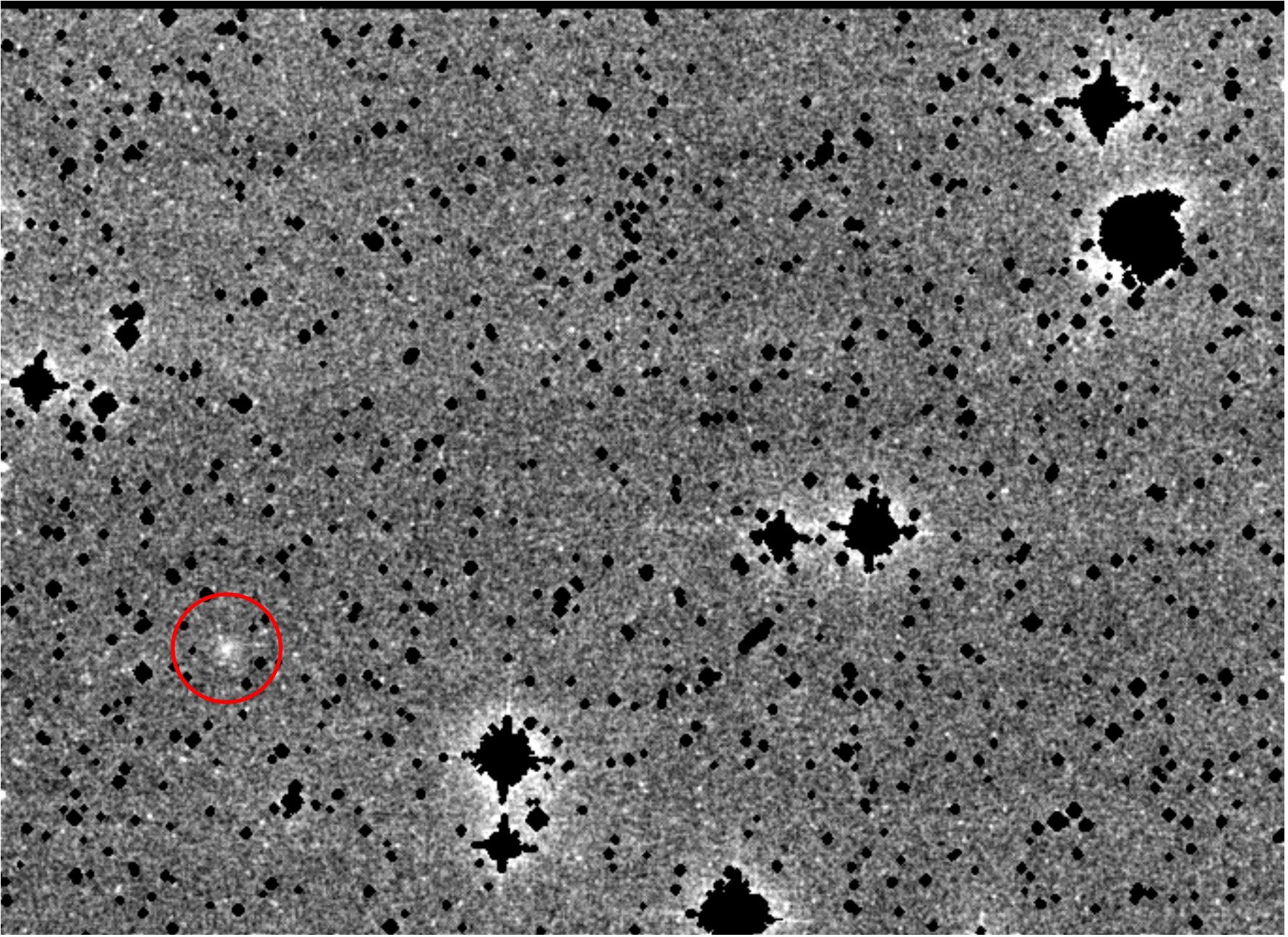}
\caption{Panels showing how the images are affected by the processing described in \S\,\ref{sec:ImRed}. 
\textbf{The top panel} shows the initial image of an example field (run 1458, filter $g$, camera column 5, field 623). 
\textbf{The middle panel} shows the same field after masking of known sources detected from the SDSS \photo\ pipeline. 
\textbf{The bottom panel} shows the field after combining the $g$, $r$ and $i$ masked images and smoothing. 
This brings out the low SB features around the brighter detected objects and gives a good representation of the challenges faced when trying to identify LSBGs. The red circle highlights the location at which an undetected LSBG becomes clearly visible after processing. The object is identified in the catalogue as LSB15283.} 
\label{fig:img_prog}
\end{figure*}

SDSS uses a drift scan mode to take images. This leads to long images spread across the sky with a width equivalent to that of the detector. For ease these images are split into `fields' of 1489 pixels by 2048 pixels, equivalent to $\sim 590\arcsec \times 810\arcsec$. The 6424 fields that were used for target selection were selected from the GAMA equatorial regions. An example of the image files taken from the SDSS is given in the top panel of Fig.~\ref{fig:img_prog}.

The images were first masked to remove high-SB objects which have been discovered using SDSS \photo\ \citep{Stoughton+02}. Any pixel classed as being associated with a detected object was masked, including stars, galaxies, cosmic ray detections, and artefacts within the image fields.
Note this is a pixel-based mask rather than a mask based on polygons 
or ellipses around detected objects, as shown in the middle panel of Fig.~\ref{fig:img_prog}. 

The data files taken from the SDSS for each field are not aligned with each other because the detectors from the different bands are not perfectly aligned in the cross-scan direction. The scan direction is in RA, for the equatorial fields, and the cross-scan direction is in DEC. 
The $g$ and $i$ images were aligned using a geometrical translation with the $r$-band coordinates as a reference, 
by up to 16 pixels difference ($\approx 6.4\arcsec$) in the $g$ band, and 5 pixels ($\approx 2\arcsec$) in the $i$ band. Once the alignment is complete all images are aligned to within one pixel ($< 0.4\arcsec$). This remaining difference is not significant enough to be of concern because we are searching for significantly extended sources.

\begin{figure}
\includegraphics[width=0.5\textwidth]{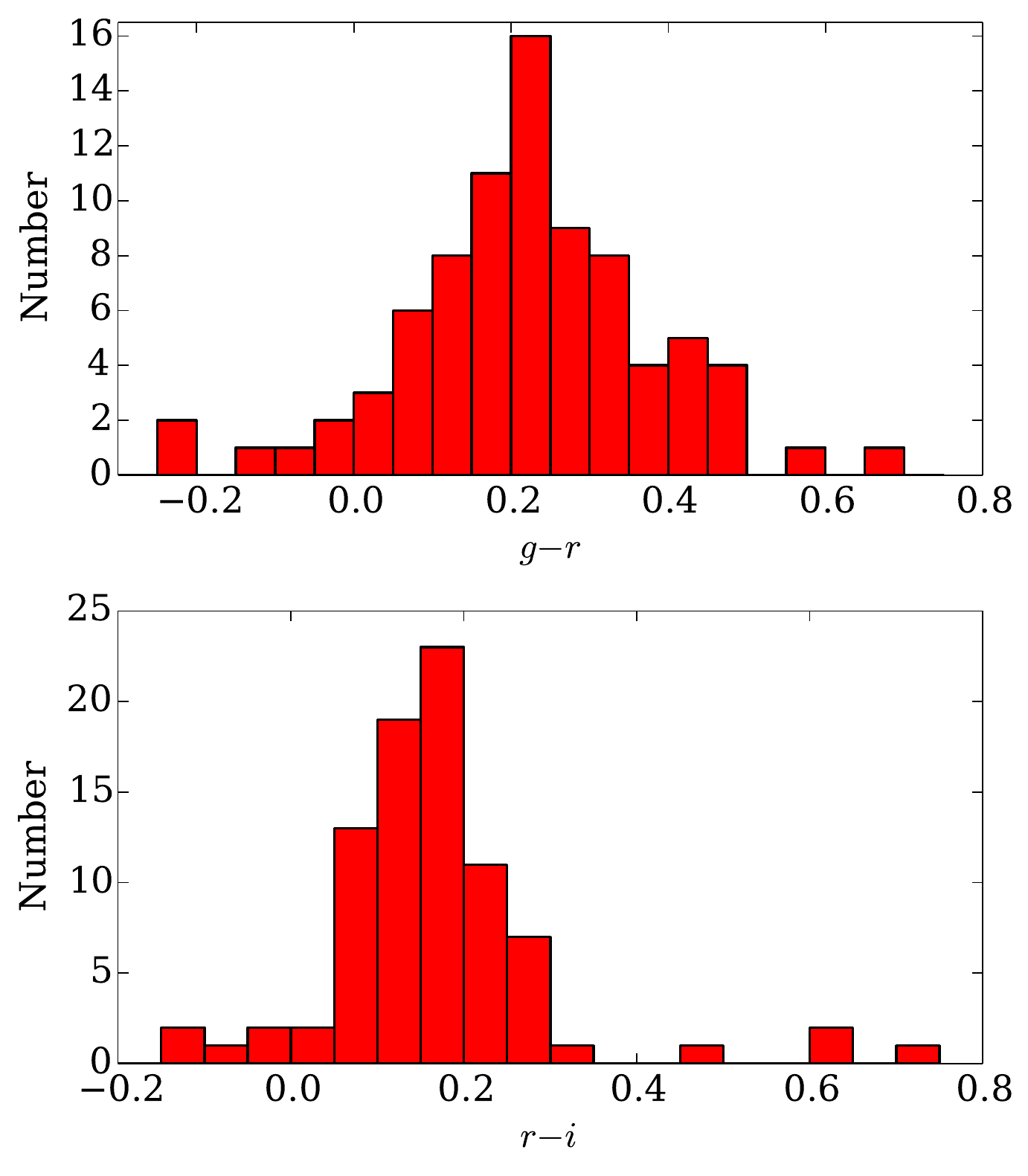}
\caption{Histograms of the $g-r$ colour of GAMA dwarf galaxies ({\bf top panel}), with a median of 0.233, 
and of the $r-i$ colour of the same galaxies ({\bf bottom panel}), with a median of 0.154. 
The galaxies used for these plots fall in the stellar mass range $10^{6}\msun < \mstar < 10^{7.5}\msun$.}
\label{fig:gricolor}
\end{figure}

LSBGs are diffuse such that the noise in any flux measurement is dominated by the sky background (with standard deviation $\sigma_{\rm sky}$). 
In order to maximise the SNR in any coadded image, the images from each filter should therefore be weighted in proportion to 
$S/\sigma_{\rm sky}^2$ where $S$ is the expected signal level of a fiducial source. 
We are interested in optimising for low-mass, star-forming (SF) dwarf galaxies. To do this, we determined the median colours
of GAMA dwarf galaxies with $10^{6}\msun  < \mstar < 10^{7.5}\msun$. These values are $g-r = 0.233$ and $r-i = 0.154$ 
with the colour distributions shown in Fig.~\ref{fig:gricolor}. 
The expected signals were then determined for a fiducial source with $(g,r,i) = (19.387,19.154,19)$,
and using the equations provided by SDSS for each field. The ranges of counts are shown 
in Table~\ref{tab:val}. 

\begin{table}
\centering
\caption{The filter, apparent magnitudes ($m$) for a fiducial source, 
the range of signals ($S$) calculated for different images, 
sky-noise values ($\sigma_{\rm sky}$), 
and weights ($\omega$) to be applied to the images before coadding.} 
\label{tab:val} 
\begin{tabular}{ c c c c c}
\hline\hline\\ [-2ex]
Filter&$m$&counts ($S$)&$\sigma_{\rm sky}$&weights $(\omega$) \\
\hline\\[-2ex] 
g&19.387 & 4283 -- 5148 & 4.37 -- 6.55 & 1\\
r&19.154 & 4169 -- 4770 & 5.29 -- 6.41 & 0.558 -- 0.921\\
i&19.000 & 3245 -- 4028 & 6.18 -- 8.01 & 0.303 -- 0.505\\
\hline  
\end{tabular} 
\end{table}

The $\sigma_{\rm sky}$ values were determined from the masked images, with the final weights relative to the $g$ band given by 
\begin{equation}
\omega  = \frac {S / \sigma_{\rm sky}^2}  {S_g / \sigma_{{\rm sky},g}^2} \mbox{~~~.}
\label{eq:weighting}
\end{equation}
The ranges in the final weights (Table~\ref{tab:val}) 
are a result of the different extinction and sky-noise conditions on the nights on which the respective fields were observed. 

The final step is to smooth the images in order to further improve the SNR. SDSS use a maximum $4 \times 4$ binning kernel for detection purposes, 
i.e.\ 16 pixels are coadded to increase the SNR. Here, we convolve an approximately circular kernel of 
diameter 7 pixels ($\approx 3\arcsec$) with each image
(in practice, the kernel is a $7 \times 7$ matrix of ones and zeros, 
for inside and outside the circle, respectively). 
This diameter is chosen as at the larger distances of interest, $\sim 100$ Mpc, we expect objects of only a
 few arcseconds on the sky \citep{Impey+1997}.
The bottom panel of Fig.~\ref{fig:img_prog} shows the effect that coadding and smoothing has on the example image. 

Now that the $g$, $r$, and $i$ bands for each field have been masked, weighted, coadded, and smoothed, using a larger kernel than was used by the \photo\ pipeline, there is increased sensitivity in the images for the discovery of field dwarf LSBGs than SDSS \photo\ was able to achieve. The next step is to develop an algorithm which can be used to detect these hitherto undetected galaxies.

\section{Detection Algorithm}
\label{sec:DetAl}

For this analysis we adopt $5\sigma$ above the background of each processed image as the detection threshold. 
Neighbouring pixels with a SNR $> 5$ are grouped into `candidate' detections. 
For each candidate detection, we record the centroid position of the grouped pixels. 
This returned about one million candidate low-SB detections. 
These need to be processed to eliminate likely false detections, and to select potentially the brightest LSBGs in terms of total flux. 

Most of candidate detections are from the excess light around bright stars and galaxies.
This excess light is evident in the bottom panel of Fig.~\ref{fig:img_prog} for a typical field.
In some fields there are also other artefacts. 
The list of sources therefore needs to be cleaned to remove these artefacts by applying a set of constraints to the detections list. 
These constraints are based on: (i) proximity to bright sources, (ii) anomalously high detection rates per field, 
and (iii) proximity to other candidate detections.
The way these constraints are applied, and the reasons for applying them, is described in the following paragraphs. 

\begin{figure}
\includegraphics[width=0.48\textwidth]{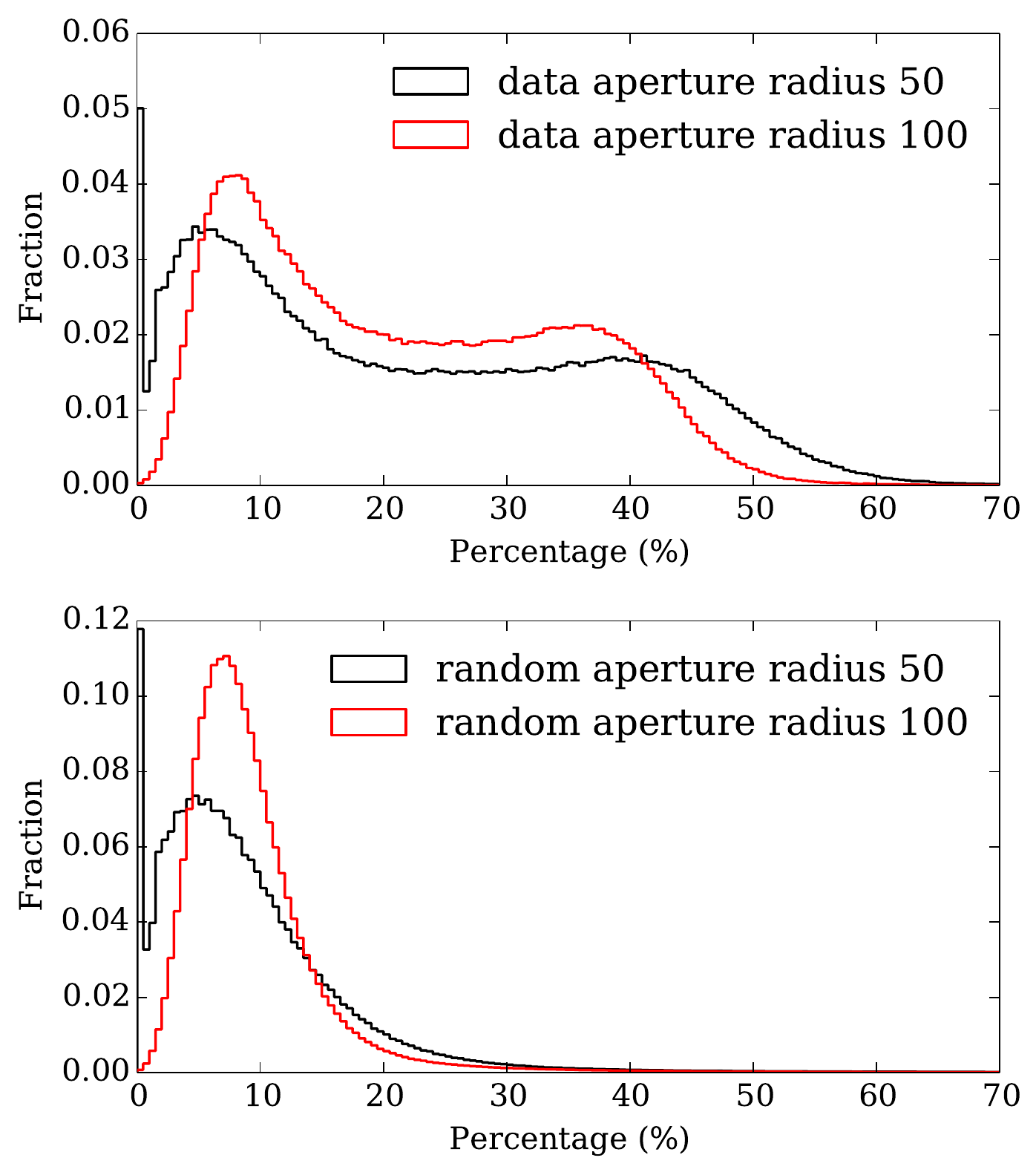}
\caption{Histograms of the percentages of masked pixels, ${\cal P}_{50}$ and ${\cal P}_{100}$, within apertures of radii 50 and 100 pixels placed 
around candidates ({\bf top panel}) and around random positions ({\bf bottom panel}). 
There is an obvious rise in the fraction of candidates that have a high percentage of masked pixels compared to randomly-placed apertures. 
This is caused by low-SB emission around bright sources. 
Candidate detections falling within this second peak can be rejected with minimal impact on the effective search area.}
\label{fig:nan_plot}
\end{figure}

In order to reject candidates that are caused by the unmasked light around bright sources, 
the percentages of masked pixels, ${\cal P}_{50}$ and ${\cal P}_{100}$, 
within circles of radii 50 and 100 pixels around each candidate were determined 
(radii of $\sim 20\arcsec$ and $40\arcsec$, respectively). 
The top panel of Fig.~\ref{fig:nan_plot} shows histograms of ${\cal P}_{50}$ and ${\cal P}_{100}$ for the candidate detections;
while the lower panel shows histograms for randomly-placed apertures, with 
1000 apertures each for ${\cal P}_{50}$ and ${\cal P}_{100}$ per field.

We would expect field LSBGs in terms of their proximity to bright sources, which are mostly stars, 
to behave more like the randomly-placed apertures. We can see from the top panel of Fig.~\ref{fig:nan_plot}
that there is an extension to large ${\cal P}$ values that is not evident for the randomly-placed apertures. 
Therefore we can reject candidates with large ${\cal P}$ values while retaining the majority of genuine LSBGs. 
The criteria used for rejection were  ${\cal P}_{50} > 15$\% and ${\cal P}_{100} > 15$\%. 
Applying this criteria to the random apertures results in 16\% of the random positions being rejected, 
therefore, this means that the effective search area is reduced to 84\% of the survey area.

\begin{figure}
\includegraphics[width=2.74cm]{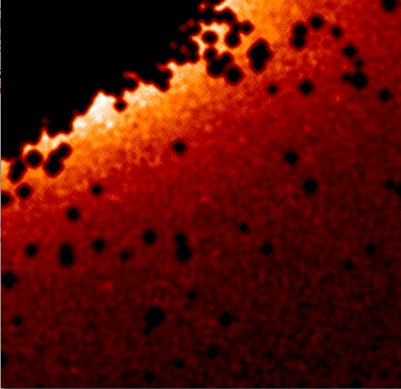}
\includegraphics[width=2.74cm]{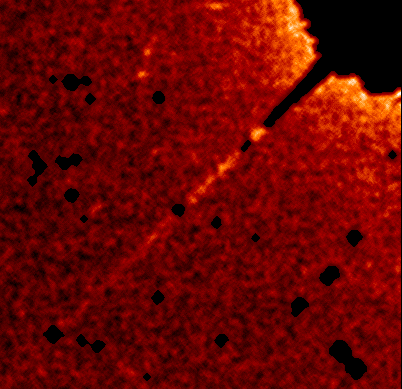}
\includegraphics[width=2.74cm]{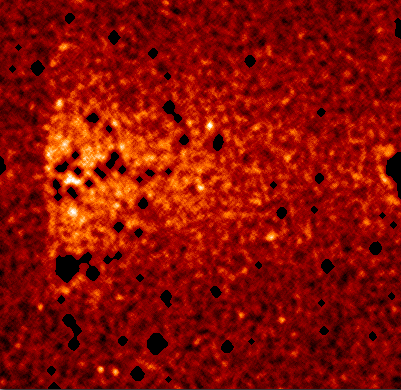}
\caption{Examples of masked-and-smoothed images demonstrating some of the issues. {\bf The left panel} shows the faint wings of the light distribution of a large galaxy. The emission from the unmasked regions causes low-SB detections. {\bf The middle panel} shows a diffraction spike; again the fainter areas of the spike are not masked. {\bf The right panel} shows an artefact caused by a bright star in a neighbouring field, giving a distinctive flat feature parallel to the edge of the image.} 
\label{fig:problems}
\end{figure}

Fig.~\ref{fig:problems} gives some examples of objects which were not removed by the masked-pixel checks carried out above, and which need to be dealt with as separate cases. The first of these are fields containing very extended wings of ultra-bright objects, making the detection of LSBGs difficult as it produces a large number of erroneous LSBG detections within the images which need to be excluded. Therefore, the constraint decided on was to reject all fields with more than 100 candidate low-SB detections within it. This affects only 22 fields, which is a small enough number as to be easily visually checked, to ensure no obvious LSBGs were rejected. 

A further issue is the detection of stray light from bright stars that are in a field adjacent to the detection. An example of this is in the right-hand image of Fig.~\ref{fig:problems}. This shows that parts of the artefact have been masked with the effect of breaking the object up into several smaller detections. In order to try to remove these objects and others like it, a constraint is applied to the images whereby all candidates that have more than 5 other candidates within a radius of $120\arcsec$ are rejected. A random sample of these objects is visually inspected to ensure that predominantly artefacts are removed from the catalogue.

One of the main reasons for conducting this study is to check the completeness of the GAMA survey, i.e., detections with $r < 19.8\,$mag, or near to this limit, are of most interest. To select potentially bright LSBGs (in total flux), two apertures of diameter $10\arcsec$ and $15\arcsec$ are placed over the objects 
and the flux measured in both on the masked images. 
We select candidates whose flux was measured to be brighter than nominally 21.3\,mag in one of the apertures for the final stage of analysis, and all others are rejected.   

This reduces the number of candidates to about 5000 detections. This sample was visually inspected. The candidates were given an integer quality rank from 0 to 2 where: 
0 means a false detection, a diffraction spike or other artefact; 
1 means a possible LSBG detection, e.g., a small object in the smoothed images with no obvious extended structure; and 
2 means a definite LSBG detection, an obvious extended source that had not been masked by SDSS.
The number of possible and definite detections after this process was 652. 
All of the removed objects are artefacts like those depicted in Fig.~\ref{fig:problems}.

\begin{figure*}
\centering
\includegraphics[width=2.75cm]{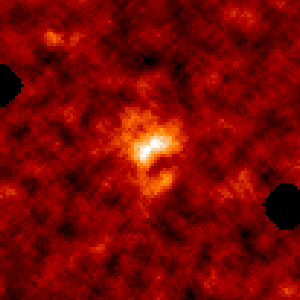}
\includegraphics[width=2.75cm]{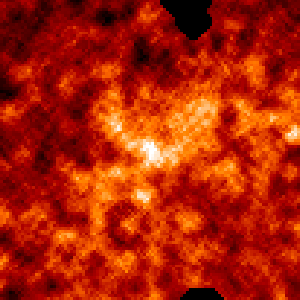}
\includegraphics[width=2.75cm]{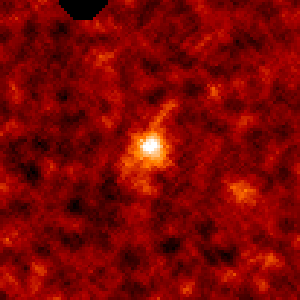}
\includegraphics[width=2.75cm]{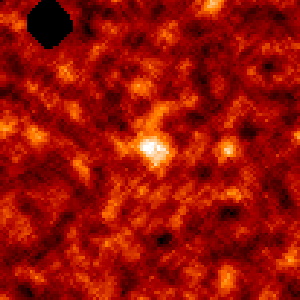}
\includegraphics[width=2.75cm]{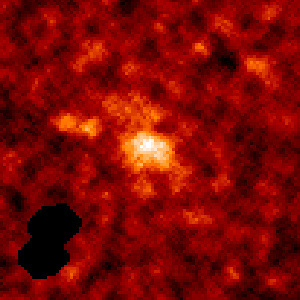}
\includegraphics[width=2.75cm]{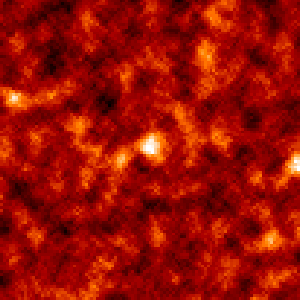}
\includegraphics[width=2.75cm]{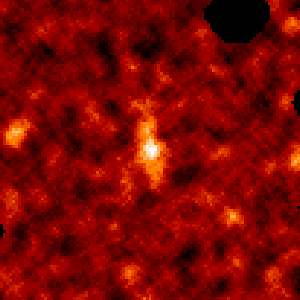}
\includegraphics[width=2.75cm]{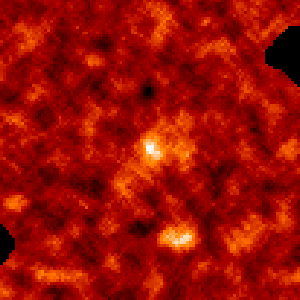}
\includegraphics[width=2.75cm]{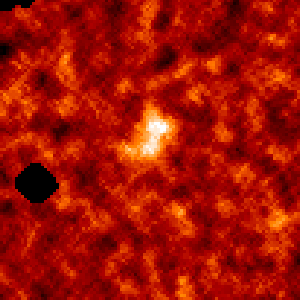}
\includegraphics[width=2.75cm]{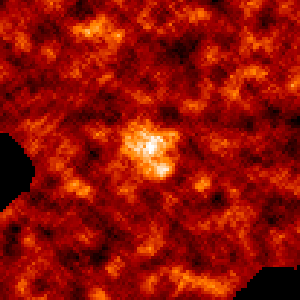}
\includegraphics[width=2.75cm]{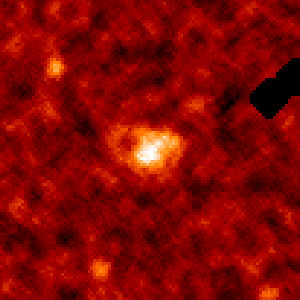}
\includegraphics[width=2.75cm]{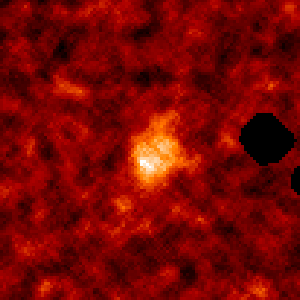}
\caption{Twelve confirmed LSBGs from the coadded, masked, and smoothed images created from SDSS $g$, $r$, and $i$ bands. 
Each candidate is positioned at the centre of the $35\arcsec \times 35\arcsec$ images. 
The detections were confirmed by deeper VIKING $Z$-band observations.}
\label{fig:objects}
\end{figure*}

The list of positions was finally visually checked against the same positions in the VIKING $Z$-band images,
which were not used not used for detection. 
These VIKING images are deeper than SDSS images for galaxies. This final check, along with eliminating duplicates of the same object, 
produced a sample of 343 LSBGs. Fig.~\ref{fig:objects} shows examples of images that were visually checked and proven to be real. 
All of the objects with rank 2 were confirmed by VIKING $Z$-band data, along with many of the rank 1 sources.\footnote{Note that 50 candidates had been assigned OBJIDs from the SDSS database. This was a concern as it was believed that all detected objects had been masked out using the SDSS \photo\ pipeline output. However, upon inspection of the SDSS flags, it was found that the objects were not detected in enough of the bands to be considered as reliable detections. We therefore kept them in the sample of 343 new detections presented in this paper.}

\section{Results}
\label{sec:Results}

\subsection{Spatial Distribution}

\begin{figure}
\includegraphics[width=8.5cm]{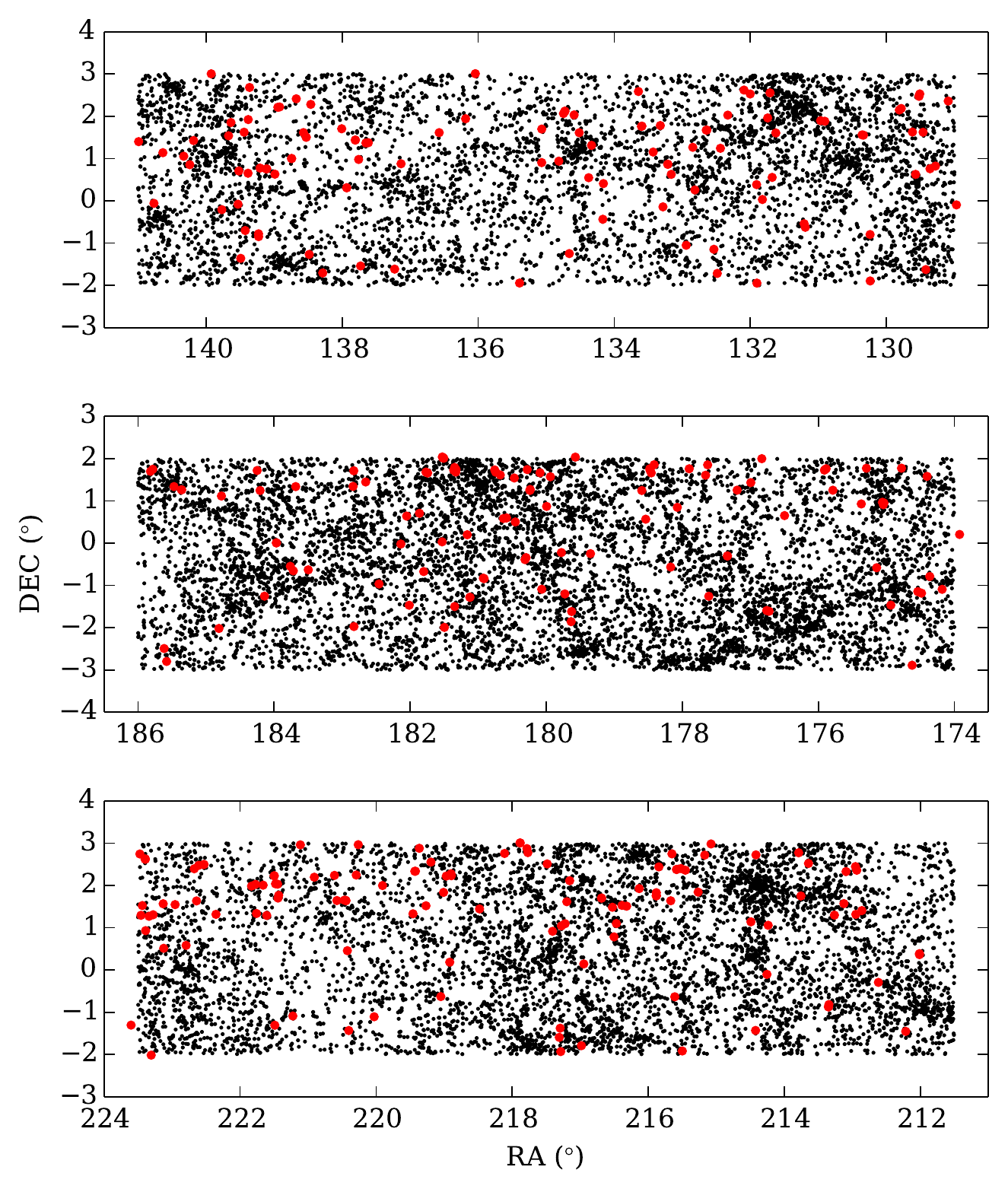}
\caption{Sky positions of the 343 LSBGs across all fields, G09, G12, and G15 (top, middle, bottom respectively; red circles). 
Also plotted are all confirmed galaxies in the GAMA survey which have known redshifts less than 0.1 (black points).}
\label{fig:pos}
\end{figure}

The distribution of large-scale structure has been consistently shown to fall into filamentary structures within a $\Lambda$CDM universe, in both simulation and through observation \citep{Press+74, Bahcall+88, Alpaslan+14}. It would be expected, therefore, that there would be some clustering of detections even for low-mass galaxies. As can be seen from Fig.~\ref{fig:pos}, the newly detected LSBGs are consistent with being associated with the $z<0.1$ large-scale structures. However, this is stated cautiously as without accurate redshift information for these objects, it is not possible to state with certainty that they are connected to these structures. 

\subsection{Magnitudes}

To select the potentially brightest LSBGs we used fixed apertures applied to the coadded-masked SDSS images. These measurements are not ideal because 
of non-optimal apertures and potential sky-subtraction uncertainties. We compute automatic apertures from the deepest band,
the VIKING $Z$ band, and measure improved matched-aperture photometry in all the SDSS and VIKING bands. 
The magnitudes were calculated using a specially designed wrapper for Source Extractor (\citealt{Bertin+96}) called \textsc{iota}.
The code is described in \cite{Driver+16} and is deployed in a similar way for this analysis.\footnote{In two cases \textsc{iota} failed to locate the 
  source in the VIKING $Z$ band because of stray light affecting the image.} 

\begin{figure}
\includegraphics[width=8.5cm]{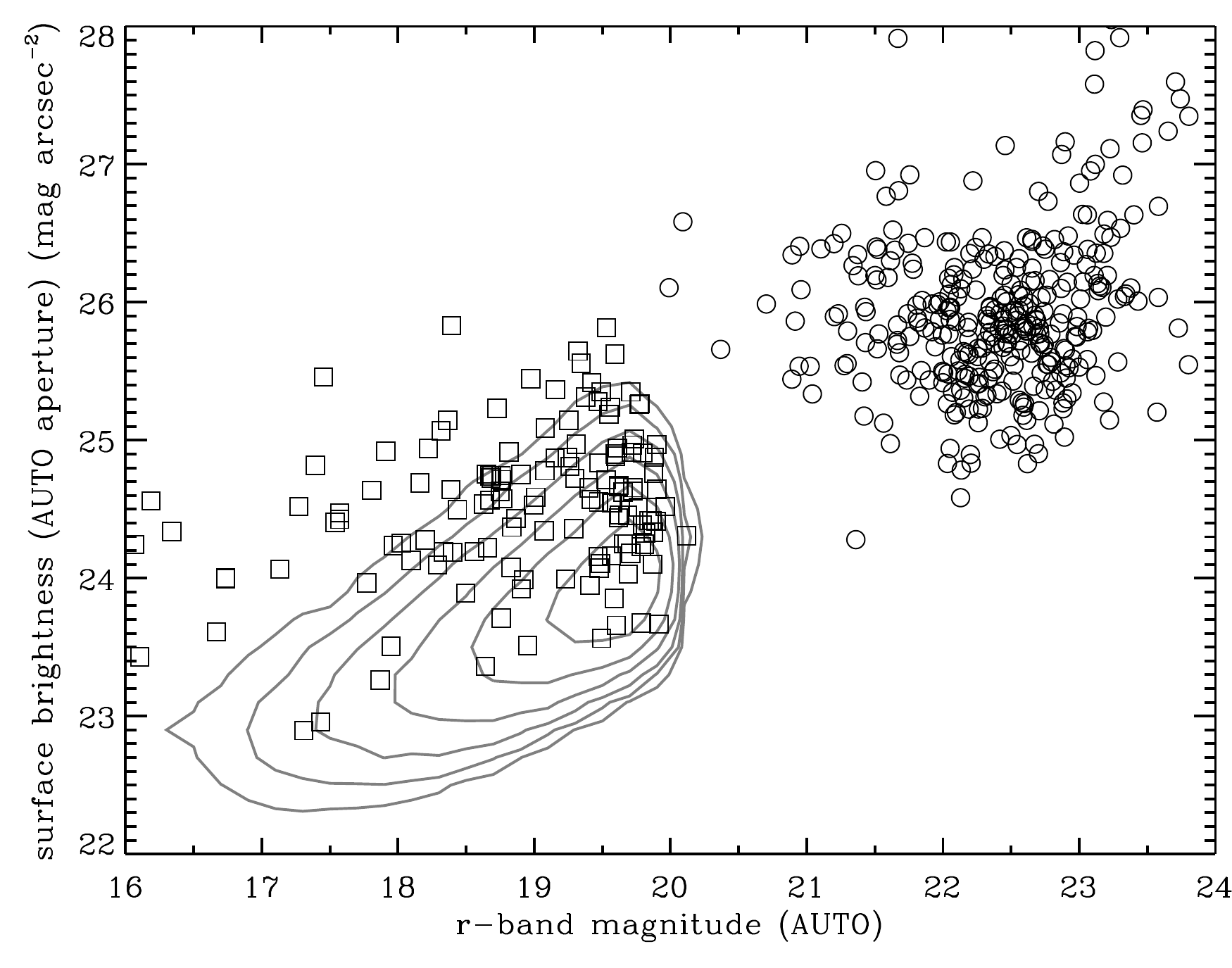}
\caption{$r$-band surface brightness versus magnitude for the GAMA main survey sample (contours), newly discovered LSBGs (circles) and the GAMA low-mass sample (squares), which were used to weight the images as described in Section~\ref{sec:ImRed}. This shows how the newly discovered systems compare to the GAMA sample and show they sit outside the main survey magnitude limit. This means that they do not affect any calculation of the GSMF using an $r < 19.8$ sample. Note the apertures for this plot were defined using the deeper VIKING $Z$ band, and that a Galactic-extinction correction
 has {\em not} been applied (sources with extinction $a_r > 0.17$ have been excluded).}
\label{fig:sbmags}
\end{figure}

Magnitudes are calculated using two apertures, a fixed aperture with a diameter of $5\arcsec$, and an \textsc{auto} aperture \citep{Bertin+96}.
The latter is used for the default magnitudes and colours, 
unless the \textsc{auto} aperture magnitude is fainter by more than 0.1 magnitudes 
than the fixed aperture, in which case, the fixed aperture magnitude is used. 
None of the $r$-band magnitudes were brighter than 19.8. 
Therefore, this LSBG sample does not have a direct effect on the calculation 
of the low-mass end of the GSMF using the GAMA main survey, 
which has an $r < 19.8$ limit. 
The surface brightness of these objects have $\mu_{r} > 24.2$\,mag\,arcsec$^{-2}$ measured within the \textsc{auto} apertures. 
Fig.~\ref{fig:sbmags} shows the distribution of SB versus magnitude for the LSBGs in comparison to GAMA samples. 
The photometry for the GAMA samples were also computed using \textsc{iota} 
with the VIKING $Z$ band as the aperture-defining band. 

\subsection{Colour Distribution}

Redshifts have not yet been determined spectroscopically for these galaxies. However, the low surface brightnesses of these objects suggest that they could be low-mass galaxies and therefore at low redshift. A useful indicator of redshift can be given by a plot of $J-K$ vs.\ $g-i$ colour, 
as shown in \cite{Baldry+10}, for $z \la 0.4$ in particular. 
The $g-i$ colour is sensitive to the 4000\AA\ break as it moves through the $g$ band, and $J-K$ is sensitive to the position of the `stellar bump'.

\begin{figure}
\includegraphics[width=8.5cm]{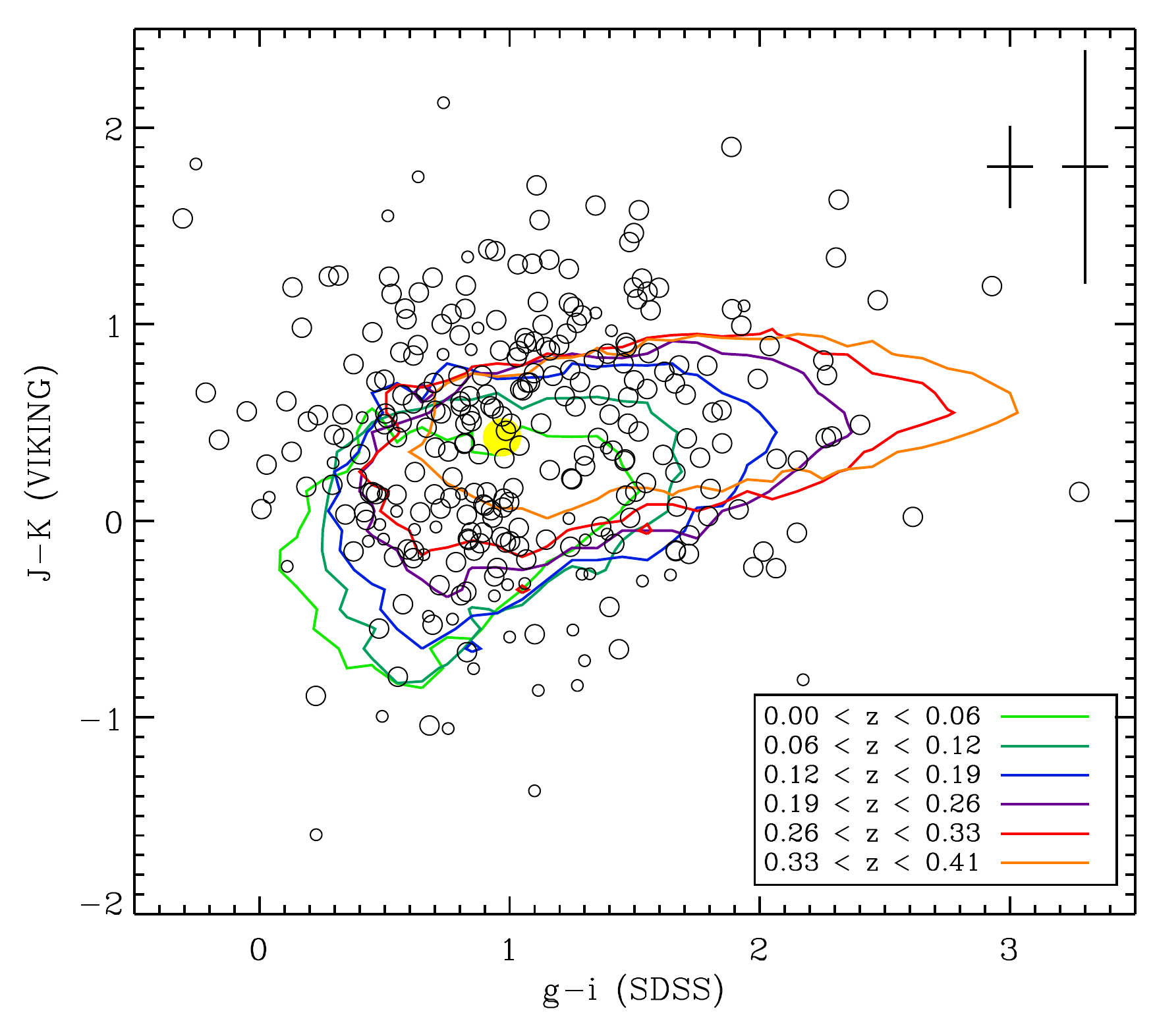}
\caption{A $g-i$ vs.\ $J-K$ colour-colour plot of all LSGBs (circles) found using the methods described in this paper. The contours show where the majority of galaxies in the GAMA fields lie in this space, split into redshift bins. This bivariate distribution is useful to determine if the object could be at low redshift \citep{Baldry+10}. The scatter for LSBGs was expected due to large uncertainties on the magnitudes. The median 1-sigma uncertainties for the large circles and 
small circles are shown in the top right of the plot (LSBGs with 1-sigma
 uncertainties larger than 1 mag are not shown). The yellow circle shows the median value of the LSBG distribution.}
\label{fig:color}
\end{figure}

Fig.~\ref{fig:color} shows the colours of the majority of the 343 sources
detected with the search algorithm, as well as distributions for GAMA redshift
samples with \textsc{auto}-aperture photometry \citep{Driver+16}. These data are
split into redshift bins and contoured to show where the peak density of each
redshift range sits on the plot.  A large proportion of the data sit around
the peaks of the lower redshift bands as shown by the median value displayed
in Fig.~\ref{fig:color}. There is scatter within this distribution, however;
this was expected as the low-SB nature of the objects means that the
uncertainties on the colours are large. Follow up is needed with spectroscopy
to determine accurate redshifts of these objects.

\subsection{Catalogue}
\label{sec:Catalogue}

Analysis of the 343 detected LSBGs were presented in the previous sections. 
These were confirmed by deeper imaging from VISTA VIKING, which
became available recently through the GAMA collaboration.  To provide a sample
that could be used, for example, to test how well detection and measurement
codes work on LSBGs, we selected a sub-sample of 57. These were all the LSBGs
measured with \textsc{iota} to have large angular extent, $a>5\arcsec$ or
$ab>15$\,arcsec$^2$, and with $r<23.5$.  This includes all twenty with
$r<21.3$.  Selected data on these 57 LSBGs are given in Table~\ref{tab:sources}.

\begin{table*}
\centering
\caption{Selected detections. $a$ and $b$ are the semi-major and semi-minor axis of the aperture fit, 
m$_{r}$ and m$_{Z}$ are the $r$-band and $Z$-band apparent magnitudes of the measured objects, 
and $\mu_{r}$ is the surface brightness within the aperture. The magnitude uncertainties from Source Extractor
are formally less than 0.1 for most of this sample; these uncertainties do not take account
of possible systematic errors in the aperture or sky subtraction. Data for this sample can be obtained
from the GAMA Panchromatic Swarp Imager at gama-psi.icrar.org.}
\label{tab:sources}
\begin{tabular}{ccrccccc}
\hline\\ [-2ex]
ID & RA        & DEC~~     & $a$      & $b$      & m$_{r}$ & m$_{Z}$ & $\mu_{r}$            \\
   & (degrees) & (degrees) & (arcsec) & (arcsec) & (mag)   & (mag)   & (mag arcsec$^{-2}$) \\
\hline\\[-2ex] 
  LSB09005     & 129.41878&  $-$1.62702  &       6.1&       4.0&      21.2&      21.1&      25.9\\
  LSB09006     & 129.45962&   1.62407  &       5.7&       4.9&      21.5&      20.6&      26.4\\
  LSB09013     & 129.80873&   2.14959  &       4.8&       4.4&      21.4&      20.7&      26.0\\
  LSB09032     & 132.43839&   1.24316  &       7.2&       5.7&      20.7&      20.7&      26.0\\
  LSB09034     & 132.53461&  $-$1.14653  &       9.2&       4.5&      21.1&      20.5&      26.4\\
  LSB09035     & 132.63810&   1.67591  &       5.8&       4.6&      23.2&      21.2&      28.0\\
  LSB09037     & 132.81347&   0.25772  &      10.5&       8.4&      20.0&      19.7&      26.1\\
  LSB09040     & 133.16144&   0.62601  &       6.8&       3.5&      21.7&      20.7&      26.4\\
  LSB09041     & 133.21152&   0.86445  &       6.3&       3.8&      23.1&      21.0&      27.8\\
  LSB09042     & 133.21306&   0.86627  &       6.8&       5.3&      21.7&      20.4&      26.8\\
  LSB09045     & 133.42739&   1.15488  &       5.5&       3.9&      20.9&      20.9&      25.4\\
  LSB09047     & 133.59885&   1.76172  &       5.7&       3.4&      23.1&      20.9&      27.6\\
  LSB09064     & 136.18438&   1.94280  &       5.3&       3.0&      21.5&      20.8&      25.8\\
  LSB09078     & 138.46061&   2.28220  &       5.1&       2.8&      22.1&      21.6&      26.2\\
  LSB09082     & 138.56844&   1.61552  &       6.1&       4.4&      21.4&      20.9&      26.2\\
  LSB09088     & 138.98769&   0.63062  &       7.9&       6.2&      20.9&      20.5&      26.4\\
  LSB09095     & 139.37987&   1.92291  &       5.0&       3.2&      21.3&      20.7&      25.5\\
  LSB09110     & 140.76613&  $-$0.05994  &       4.8&       4.1&      21.4&      21.4&      25.9\\
  LSB09111     & 140.99066&   1.40064  &       5.8&       5.0&      21.6&      21.1&      26.5\\
  LSB12115     & 174.40137&   1.57954  &       9.2&       5.2&      20.9&      20.6&      26.3\\
  LSB12119     & 174.77714&   1.77343  &       5.4&       3.7&      21.0&      21.2&      25.5\\
  LSB12133     & 176.83034&   1.99869  &      11.0&       9.1&      21.7&      18.9&      27.9\\
  LSB12143     & 178.16978&  $-$0.56624  &       4.8&       3.2&      22.9&      21.0&      27.1\\
  LSB12153     & 179.63320&  $-$1.85774  &       5.1&       4.3&      20.9&      20.7&      25.5\\
  LSB12156     & 179.93796&   1.57097  &       5.0&       2.9&      21.8&      21.1&      26.0\\
  LSB12159     & 180.09084&   1.66413  &       6.2&       4.8&      21.3&      20.5&      26.3\\
  LSB12167     & 180.46868&   1.54137  &       6.1&       3.8&      22.2&      20.5&      26.9\\
  LSB12168     & 180.58229&   0.59635  &       7.3&       3.9&      21.5&      20.6&      26.4\\
  LSB12183     & 181.34540&   1.79183  &      11.9&      10.6&      20.1&      19.7&      26.6\\
  LSB12187     & 181.50472&   2.00895  &       4.3&       3.8&      22.9&      22.0&      27.2\\
  LSB12196     & 182.01351&  $-$1.46978  &       7.8&       4.7&      21.8&      20.5&      26.9\\
  LSB12200     & 182.65347&   1.44755  &       5.5&       3.9&      21.6&      21.1&      26.2\\
  LSB12214     & 184.81006&  $-$2.01908  &       5.2&       4.6&      21.6&      20.8&      26.3\\
  LSB12218     & 185.57948&  $-$2.79934  &       7.0&       6.9&      21.5&      19.8&      27.0\\
  LSB12221     & 185.81628&   1.69855  &       4.9&       3.8&      22.0&      20.9&      26.4\\
  LSB15232     & 213.26299&   1.29825  &       5.2&       3.8&      21.3&      20.8&      25.8\\
  LSB15237     & 213.75583&   1.75108  &       5.8&       4.3&      21.6&      21.2&      26.4\\
  LSB15239     & 214.23705&   1.05752  &       5.0&       3.9&      21.8&      21.1&      26.2\\
  LSB15244     & 215.07665&   2.98220  &       9.2&       4.1&      21.6&      20.6&      26.8\\
  LSB15245     & 215.16892&   2.71834  &       6.3&       3.8&      21.5&      20.2&      26.2\\
  LSB15249     & 215.51984&   2.40010  &       5.6&       4.0&      21.9&      20.9&      26.5\\
  LSB15250     & 215.58328&   2.37787  &       8.3&       4.4&      21.0&      20.7&      26.1\\
  LSB15251     & 215.58316&   2.37885  &       5.2&       2.8&      22.2&      21.4&      26.4\\
  LSB15267     & 216.68654&   1.69957  &       5.3&       4.5&      22.5&      20.7&      27.1\\
  LSB15274     & 217.28680&  $-$1.93128  &       5.3&       3.1&      21.3&      22.1&      25.6\\
  LSB15280     & 217.78322&   2.87196  &       5.4&       3.8&      21.8&      21.1&      26.3\\
  LSB15283     & 218.11156&   2.75746  &       5.7&       2.9&      21.0&      20.7&      25.3\\
  LSB15284     & 218.47939&   1.44470  &       5.1&       4.7&      21.2&      20.5&      25.9\\
  LSB15286     & 218.89958&   2.27586  &       5.8&       3.9&      23.3&      21.3&      27.9\\
  LSB15297     & 219.90555&   1.99790  &       6.8&       4.5&      20.9&      20.9&      25.9\\
  LSB15305     & 220.46707&   1.65167  &       8.7&       4.6&      21.3&      20.3&      26.5\\
  LSB15307     & 220.61205&   2.23853  &       6.5&       4.8&      21.4&      20.6&      26.3\\
  LSB15308     & 220.90856&   2.19508  &       5.2&       3.1&      21.4&      21.1&      25.7\\
  LSB15326     & 222.60628&   2.47710  &       8.0&       4.9&      21.2&      20.4&      26.4\\
  LSB15329     & 222.79291&   0.58235  &       4.8&       3.8&      22.1&      20.8&      26.4\\
  LSB15330     & 222.95473&   1.54644  &       4.8&       4.8&      21.5&      21.2&      26.2\\
  LSB15336     & 223.38822&   0.93069  &       8.4&       4.9&      20.4&      20.3&      25.7\\
\hline 
\end{tabular}
\end{table*}

\section{Summary and Conclusions}
\label{sec:sum+con}

This work attempts to answer a simple question: are there any LSBGs hidden within the GAMA equatorial regions that could contribute to the low-mass end of the GAMA GSMF? Using images from the SDSS, and a specially developed algorithm to process the images and detect the objects it was discovered that whilst there are LSBGs, they do not meet the required magnitude cut of $r < 19.8\,$mag. Therefore they do not affect the GAMA GSMF at low masses as presented in \cite{Baldry+2012}.
If they are low-mass galaxies, they could be significant for any attempt to measure further down the GSMF
using a deeper sample such as from the Wide Area VISTA Extragalactic Survey (WAVES; \citealt{DriverWave+16}). 

The algorithm created consisted of several parts: the weighting and coadding of the images, masking, and smoothing to bring out any hidden objects within the images. A cut of $5\sigma$ was then applied to the images to identify any pixels with a high enough SNR to be considered a detection. After clumping the detected pixels into candidate objects, a set of constraints were applied to the these objects. This removed most of them as erroneous detections such as from extended wings of bright stars and galaxies that were not masked out, and from stray light from bright stars in neighbouring fields. After a final comparison to VIKING $Z$ band, 343 new galaxy detections were confirmed.

The magnitudes and surface brightnesses of the final sample were determined primarily using an \textsc{auto} aperture. The majority of objects were consistent with being at low redshift, $z<0.2$, when comparing a $J-K$ vs $g-i$ plot of all candidates to the GAMA main survey (Fig.~\ref{fig:color}). This plot is a good proxy for photometric redshift and can give a visual indication of whether the objects are at low or high redshift. Only a minority are likely to be in our cosmological
neighbourhood within 100\,Mpc, however, it should be noted that the uncertainties in the colours are 
probably underestimated because of the difficulty in measuring accurate photometry of LSBGs.

Fig.~\ref{fig:sbmags} shows how the newly discovered sample compare to the main GAMA survey in terms of surface brightness and magnitude. It is clear that the systems discovered in this work are too faint to be included in any calculations of the GSMF using the GAMA main survey limit. 
However, the LSBG catalogue can be used in future studies as a test sample for deeper imaging in the same regions. 
Source detection software run on deeper imaging such as KiDS and VIKING should readily detect these galaxies, however, this is by no means a given as
errors in sky subtraction and/or flat fielding can cause problems in identifying and characterising low-SB features and galaxies. 
In future, we plan to use a source extraction run on the VIKING-$Z$ band mosaics matched to the GAMA redshifts to improve 
on estimates of the low-mass end of GSMF. 

\section*{Acknowledgments}
GAMA is a joint European-Australasian project based around a spectroscopic campaign using the Anglo-Australian Telescope. The GAMA input catalogue is based on data taken from the Sloan Digital Sky Survey and the UKIRT Infrared Deep Sky Survey. Complementary imaging of the GAMA regions is being obtained by a number of independent survey programmes including GALEX MIS, VST KiDS, VISTA VIKING, WISE, Herschel-ATLAS, GMRT and ASKAP providing UV to radio coverage. GAMA is funded by the STFC (UK), the ARC (Australia), the AAO, and the participating institutions. The GAMA website is http://www.gama-survey.org/ . 

%\bibliographystyle{mn2e-williams}

%\bibliography{References}

\label{lastpage}

\end{document}